\documentstyle[12pt]{article}
\def\be{\begin{equation}}
\def\ee{\end{equation}}
\def\bea{\begin{eqnarray}}
\def\eea{\end{eqnarray}}

\author{Hans - J\"urgen Schmidt}

\title{Inhomogeneous Cosmological Models Containing Homogeneous 
Inner Hypersurface Geometry. Changes of the {\sc  Bianchi}  Type. }

\date{}
\begin{document}
\maketitle

\centerline{Universit\"at Potsdam, Institut f\"ur Mathematik, Am
Neuen Palais 10} 
 \centerline{D-14469~Potsdam, Germany,  E-mail:
 hjschmi@rz.uni-potsdam.de}

\begin{abstract}
There are
 investigated such cosmological models which instead of the usual spatial
 homogeneity property only fulfil
 the condition that in a certain synchronized system of reference all
spacelike 
sections    $t$ = const. are homogeneous manifolds.
This allows time-dependent changes of the BIANCHI type. 
Discussing differential-geometrical theorems it is shown which of them are 
permitted.  Besides the trivial case of changing into type I
 there exist some possible changes between other types. However, physical
reasons
 like energy inequalities partially exclude them.

\bigskip

\noindent
Es werden kosmologische Modelle betrachtet,
 die anstelle der   \"ublichen r\"aumlichen Homogenit\"atseigenschaft 
nur die Bedingung   erf\"ullen, da\ss \   in einem gewissen synchronisierten 
Bezugssystem alle  Raumschnitte $t$ = const. 
homogene Mannigfaltigkeiten sind. 
Neben dem trivialen Fall des Wechsels zu Typ I
 gibt es einige m\"ogliche \"Anderungen zwischen anderen Typen. 
Physikalische Gr\"unde wie Energieungleichungen schlie\ss{}en 
diese jedoch teilweise aus.
\end{abstract}

\section{Introduction}

Recently,   besides the   known BIANCHI models,
 there are investigated certain classes of inhomogeneous cosmological
 models.   This is done to
get a better representation of the really existent inhomogeneities, 
cf. e.g. BERGMANN (1981), CARMELI (1980),
 COLLINS (1981), SPERO (1978), SZEKERES (1975) and WAINWRIGHT (1981). 
We consider,   similar as in   COLLINS  (1981), 
such inhomogeneous  models $V_4$
 which in a certain synchronized system of 
 reference possess homogeneous  sections $t$ = const., called 
$V_3(t)$ \footnote{This is   analogous  to  the generalization
 of the concept of spherical symmetry in KRASINSKI (1980).}.   
In the  present paper we especially investigate which
 time-dependent changes   of the BIANCHI type are possible. Thereby 
we impose, besides the twice continuous differentiability, 
 a physically reasonable condition: 
the energy inequality,   $T_{00} \ge \vert T_{\alpha \beta}\vert $, 
holds in each LORENTZ frame.

Under this point  of view, we make some globally topological remarks: 
Under the physical condition the topology  of the sections $V_3(t)$ is, 
according to LEE (1978), independent of $t$.  Hence, the 
KANTOWSKI-SACHS models,  with underlying topology 
${\rm S}^2 \times {\rm R}$ or ${\rm S}^2 \times {\rm S}^1$  and the 
models of BIANCHI type IX with 
underlying topology ${\rm S}^3$  or continuous images of it as 
SO(3) may not change, because all other types are represented by
the R$^3$-topology, factorized with reference to a discrete subgroup of
 the group of motions. But  the remaining types can all be 
represented in R$^3$-topology itself; therefore we do not get any
further global restrictions. All homogeneous  models, the  above mentioned
KANTOWSKI-SACHS model being excluded, possess simply transitive 
groups (at least subgroups) of motion. 
 Hence we do not specialize if we deal in the following only locally with
simply-transitive groups of motions.\footnote{We consequently do not 
consider here trivial changes of the  BIANCHI  type, e. g.  from
type III to type VIII by means of an intermediate on which a group of motions
 possessing  transitive subgroups  of both types acts.}

To begin with, we consider the easily tractable case of a change to type I:
  For each  type M there  one can find a manifold $V_4$
 such that for each $t \le  0$  the section $V_3(t)$
 is flat and for each $t > 0$
it belongs to type M. Indeed, one has simply to use for 
all $V_3(t)$  such representatives of type M
that their curvature vanishes 
 as $t \to 0_+$.\footnote{Choosing exponentially decreasing curvature 
one can obtain an arbitrarily high differentiable
class for the metric, e. g.
$$
     ds^2 =  -dt^2 + dr^2 +  h^2 \cdot (d\psi^2 + \sin^2 \psi d\phi^2) \, ,
$$
where $h =  r$
 for $t \le 0$
 and 
$$
 h = \exp (t^{-2}) \cdot \sinh^2 \left( r \cdot \exp(-t^{-2}) \right)
$$
else. This is a $C^\infty$-metric whose $V_3(t)$
belong to type I for $t \le 0$  and to type
V for $t > 0$.
 (In $t =  0$, of course, it cannot be an analytical  one.) The limiting
slice belongs 
by continuity causes necessarily to type I.}
 Applying this fact twice it becomes obvious that by the
help of a flat intermediate (of finite extension) or only by a single
 flat slice all BIANCHI types
can be matched together. However, if  one does not want to use 
such a flat intermediate  the
transitions of one BIANCHI type  to another become a non-trivial 
problem. It is shown as well from the purely differential-geometrical
 as from the  physical points of view (sections 3 and 4
resp.)  which  types can be matched  together immediately 
 without a  flat  intermediate. To this end
we collect the following preliminaries.

\section{Spaces possessing homogeneous slices}

As one knows, in cosmology the homogeneity principle is expressed by the 
fact that to a
space-time $V_4$
 there exists a group of  motions acting transitively on the spacelike 
hypersurfaces $V_3(t)$  of a slicing of $V_4$. Then the metric
is given by
\be
ds^2 = - dt^2 + g_{ab}(t) \omega^a  \omega^b
\ee
where the $g_{ab}(t)$  are  positively definite and $\omega^a$ 
are the 
basic 1-forms corresponding  to a certain
BIANCHI  type.  If  the   $\omega^a$    are  related to a 
holonomic basis $x^i$, using type-dependent functions
$ A^a_i(x^j)   $, one 
can write:
\be
\omega^a =  A^a_i dx^i    \, .
\ee
For the spaces considered here we have however: 
there is a synchronized system  of reference
such that the slices $t= $  const. are homogeneous
 spaces $V_3(t)$. In this system of reference the
metric is given by
\be
     ds^2 = - dt^2 + g_{ij} dx^i dx^j 
\ee
where   $g_{ij}(x^i,t)$   are  twice continuously differentiable and 
homogeneous  for constant $t$. Hence
it is a generalization of (1) and (2). This is a genuine generalization 
because there is only
posed the condition $g_{0\alpha} = - \delta_{0\alpha}$
 on the composition of  the homogeneous slices 
$V_3(t)$  to a $V_4$.
 This means in each  slice only the first fundamental form
 $g_{ij}$  is homogeneous;  but in the contrary 
to homogeneous models, the second fundamental form 
$\Gamma_{0ab}$  need not have this property.
Therefore, also the curvature scalar   $^{(4)}R$   
(and with it the distribution of matter) need not be
constant within a $V_3(t)$.  Now let $t$ be fixed. In $V_3(t)$
 one can find then  coordinates     $ x^i_t (x^j,t)  $
such that according to (1)  and (2) the inner metric gets the form
\be
 g_{ab}(t) \omega^a_t  \omega^b_t   \qquad {\rm where} \qquad
\omega^a_t =  A^a_i (x^i_t) dx^i_t \, .
\ee
Transforming this into the original coordinates $x^ i$
 one obtains for the metric of the full $V_4$
\be
g_{0\alpha} = - \delta_{0\alpha} \, , \qquad 
g_{ij} = g_{ab}(t) \, A^a_k (x^i_t) \ A^b_l (x^i_t) \,  
x^k_{t,i} \,   x^l_{t,j} \, .
\ee
If  the BIANCHI  type changes with  time one has to take 
such an ansatz for each interval of
constant type separately; between them  one has to secure
 a $C^2$-joining.\footnote{ An example of such an inhomogeneous model 
 is  (cf. ELLIS 1967)
$$  
   ds^2 =   -dt^2  + t^{-2/3} \left[ t + C(x) \right]^2 
dx^2 +   t^{4/3} (dy^2 + dz^2) \,  .
$$
The  slices $t =$ const. are  flat, 
but only  for constant $C$ it belongs to BIANCHI type I.}

\section{Continuous changes of the BIANCHI type}

Using the usual homogeneity property the
 same group of motions acts on each slice $V_3(t)$.
Hence the BIANCHI  type is independent of time
 by definition. But this fails to be the case for
the spaces considered here. However, we can deduce the following:
 completing $\partial/\partial t$ to an
anholonomic basis which is connected with KILLING vectors 
in each $V_3(t)$  one obtains for the
structure constants\footnote{ These structure constants  are  calculated 
 as follows: Without loss of generality  let $x^i_t(0,0,0,t)=0$.
At this point 
$\partial/\partial x^i_t$
 and 
$\partial/\partial x^i$ 
 are  taken as initial values for KILLING vectors within $V_3(t)$. 
The  structure constants obtained by these KILLING vectors are
 denoted by   
$  \bar  C^i_{jk} (t) $
 and $ C^i_{jk} (t) $
resp.  where
$  \bar  C^i_{jk} (t) $ 
 are the canonical ones. Between them it holds at $(0,0,0,t)$:
\be
x^i_{t,j} C^j_{kl} =  \bar  C^i_{mn}
 x^m_{t,k} x^n_{t,l} \, . 
\ee
}
 associated to the commutators of  the basis:
\be
C^0_{\alpha \beta} =0 \, ; \quad
 C^i_{jk}(t)  \quad  {\rm depends \  continuously \  on \   time}.        
\ee
      (Using only the usual
canonical structure constants then of
 course no type is changeable continuously.)

To answer the question which BIANCHI types may change 
continuously we consider all sets of
structure constants 
 $ C^i_{jk}  $
 being antisymmetric in $jk$ and fulfilling
 the JACOBI  identity. Let   $C_{\rm S}$  be
such a set belonging to type S. Then we have
 according to (7): Type R is changeable
continuously into type S, symbolically expressed by the 
validity of    R $\to$  S, if  and only if to each
   $C_{\rm S}$ 
there exists a sequence 
$ C_{\rm R}^{(n)}  $
 such  that holds,
\be
\lim_{n \to \infty}  C_{\rm R}^{(n)}   =  C_{\rm S} \, .
\ee
The limit has to be understood componentwise.

It holds that R $\to$   S  if and only if there are
 a     $C_{\rm S}$  and a sequence 
 $ C_{\rm R}^{(n)}  $  such that (8)  is fulfilled.
The equivalence of both statements is shown by 
means of simultaneous rotations of the basis.
In practice one takes as 
 $C_{\rm S}$ 
 the canonical structure constants and investigates for which types
R there can be found corresponding sequences 
$ C_{\rm R}^{(n)}  $:
the components of    $C_{\rm S}$  
 are subjected to a perturbation not exceeding $\epsilon$
 and their BIANCHI types
are  calculated.   Finally one  looks which types appear 
for all   $\epsilon > 0$.   Thereby one profits, e. g.,
by the   statement that the dimension of 
 the image space of the LIE    algebra  (0  for type I,
\dots,  3 for types VIII and IX)  cannot increase during such changes.

One obtains the  following diagram. 
The validity of R $\to$   S and S $\to$  T also imply the validity 
of R  $\to$ T, hence  the   diagram must be  continued transitively. 
The statement 
 VI$_\infty \  \to $ IV expresses the
 facts that there exists a sequence 
$$
C^{(n)}_{{\rm VI}_h} \to C_{\rm IV}
$$
and that in each
such sequence the parameter $h$  must necessarily  tend to  infinity.

Further VI$_h \  \to$  II holds for all $h$.
 Both statements  hold analogously for type  VII$_h$. In
MACCALLUM (1971)  it is shown a similar diagram but
 there it is only investigated 
which  changes appear if some of the canonical structure
 constants are  vanishing.  So e. g., the different
transitions from  type VI$_h$  to types II and IV are not 
contained in it.

\bigskip

\centerline{Abb. 1}

\bigskip

To complete the answer to the question 
posed above it must be added: to each transition shown in the diagram
 there one can indeed find a $V_4$  in  which it is realized.\footnote{Let
 $f(t)$  be a $C^\infty$-function
 with $f(t) = 0$  for $t \le 0$  and $f(t) > 0$ else. Then, e. g., 
$$
ds^2 = -dt^2 + dx^2
+ e^{2x} dy^2 + e^{2x}(dz + x f(t)  dy)^2
$$
 is a  $C^\infty$-metric  whose slices $V_3(t)$ belong to type V  
and IV for $ t \le
0$ and $t > 0$ resp. 
Presumably it is typical that in  a neighbonrbood of $t =0$  
the curvature becomes 
singular at spatial infinity.}
 Actually, for  the transition R $\to$  S  the limiting slice
 be1ongs  necessarily to type S.

\section{Physical conditions}

To obtain physically reasonable space-times
 one has at least to secure the validity of an energy
inequality, 
e.g. $T_{00} \ge \vert T_{\alpha \beta} \vert $
 in each LORENTZ frame. Without this requirement
 the transition II $\to$ I
 is possible. We prove that the requirement $T_{00} \ge  0$ 
alone is sufficient to forbid this transition.
To this end let $V_4$  be a manifold which in a certain
 synchronized system of   reference possesses
a flat slice $V_3(0)$  and for all $t > 0$ type II-slices $V_3(t)$.
 Using the notations of   footnote 5 we  have: 
$ \bar  C^1_{23} = -   \bar  C^1_{32} =1  $
 are the only non-vanishing canonical 
structure constants of  type II. With  the
exception $A_2^1= - x^3$
    we have $A^a_i = \delta^a_i$.
 By the help of (6)  one can calculate the structure 
constants $ C^i_{jk}(t)$. The
flat-slice condition is  equivalent to
\be
\lim_{t \to 0} 
C^i_{jk} (t) =0 \, .
\ee
First we consider the special case
$x^i_t = a^i(t) \cdot x^i$
  (no sum), i. e. extensions of  the coordinate axes. It holds 
 $x^i_{t,j} = \delta ^i_j a^i $
 (no sum). The only
 non-vanishing     $C^i_{jk} (t)$
 are 
$ C^1_{23} = -  C^1_{32} =
   a(t) $, 
where $ a = a^2 \cdot  a^3/a^1$.
 Then (9) reads $\lim_{t \to 0}
 a(t) = 0$. The coefficients $g_{ab}(t)$ have to be chosen such 
that  the $g_{ij}$, according  to (5), remain positive definite
and twice continuously differentiable. Let $^{(3)}R$
 be the scalar curvature within
 the slices.    $^{(3)}R > 0$
appears only in type IX and in the KANTOWSKI-SACHS models. 
But changes from these
types are just the cases already excluded by global considerations.
 Inserting      $g_{ij}$ 
 into the
EINSTEIN equation by means of  GAUSS-CODAZZI  theorem 
one obtains
\be
\kappa T_{00} = R_{00} - \frac{1}{2} g_{00} \,  {}^{(4)}R = \frac{1}{2} 
 \,  {}^{(3)}R + \frac{1}{4} g^{-1} \cdot H \, , \quad 
 g = \det g_{ij} \, , 
\ee
$^{(3)}R$
 being the scalar curvature within the slices, hence
\bea
^{(3)}R \le 0 \ {\rm and } \nonumber \\
 H =
g_{11}[g_{22,0} g_{33,0} - (g_{23,0})^2] 
- 2 \cdot g_{12} [ g_{33,0} g_{12,0}
- g_{13,0} g_{23,0} ]
+ {\rm cyclic \  perm}.
\eea
In our case $H$  is a quadratic  polynomial in $x^3$ whose
 quadratic coefficient  reads
$$
- g_{11} [g_{11}g_{33} - (g_{13})^2] \cdot
 \left[ \frac{\partial}{\partial t} a(t) 
\right]^2 \, . 
$$
Hence, for sufficiently  large  
values $x^3$  and  values $t$  with 
$\partial/\partial t \, a(t) \ne 0$
we have $H <0$,  and
therefore $T_{00}<0$.

Secondly we  hint at another special case, namely rotations of the
 coordinate axes  against each other, e. g.
$$
x^1_t = x^1 \cos \omega + x^2 \sin \omega \, , \quad
x^2_t = x^2 \cos \omega - x^1  \sin \omega \, , \quad
x^3_t = x^3 \, ,   \quad \omega =
\omega
(t) \, .
$$
There one obtains in analogy negative $T_{00}$. 
Concerning the general case  we  have: loosely
speaking, each diffeomorphism  is  a composite 
of such extensions and rotations. 
Hence, for
each II $\to$ I-transition one would obtain points with
 negative  $T_{00}$.

Hence, the energy condition is 
a genuine restriction to the possible transitions of  the  BIANCHI
types.  Concerning the other 
transitions we remark: equations (10) and (11) keep valid, and the
remaining work is to 
examine the signs of the corresponding expressions $ H$.
 Presumably one  always obtains points 
with negative $H$,  i.e., also under this weakened homogeneity
presumptions the BIANCHI types 
of the space-like hypersurfaces at different distances
 from the singularity must coincide.

\noindent 
I am grateful to Prof. H.-J. TREDER and Dr.
 H.-H. v. BORZESZKOWSKI  for helpful
 discussions.

\section*{References}

\noindent 
BERGMANN, O.:  1981, Phys. Lett. A 82, 383.

\noindent 
CARMELI,  M.  Ch. CHARACH: 1980, Phys. Lett. A 75, 333.

\noindent 
COLLINS,  C.  B.,  D. A. SZAFRON: 1981,  J. Math. Phys. 22, 543.

\noindent 
ELLIS,  G.:  1967,  J. Math. Phys. 8, 1171.

\noindent 
KRASINSKI, A.:  1980, GR 9-Abstracts, Jena, 44.

\noindent 
LEE,  C. W.:  1978, Proc. R. Soc. Lond. A 364, 295 and references cited there.

\noindent 
MACCALLUM,  M.:  1971, Commun. Math. Phys. 20, 57.

\noindent 
SPERO, A.,   R.  BAIERLEIN: 1978, J. Math. Phys. 19, 1324.

\noindent 
SZEKERES, P.:  1975, Commun. Math. Phys. 41, 55.

\noindent 
WAINWRIGHT,  J. : 1981, J.  Phys. Lond. A 14, 1131.

\bigskip

\noindent
With 1 figure (Received 1981 November 9)

\medskip

\noindent 
{\small {In this reprint (done with the 
kind permission of the copyright owner) 
we removed only obvious misprints of the original, which
was published in Astronomische Nachrichten:   
 Astron. Nachr. {\bf 303} (1982) Nr. 4, pages 227 - 230;  
  Author's address that time:  
Zentralinstitut f\"ur  Astrophysik der AdW der DDR, 
1502 Potsdam--Babelsberg, R.-Luxemburg-Str. 17a.}}

\newpage

\setlength{\unitlength}{3cm}
\begin{picture}(4.5,6.4) 
\thicklines
\put(4,1.5){\vector(-1,0){1.8}} 

\put(4,0){\vector(-1,0){1.8}}

\put(4.2,1.3){\vector(0,-1){1.1}} 

\put(4.2,2.8){\vector(0,-1){1.1}} 

\put(3.45,2.96){\vector(-1,-1){1.2}} 

\put(2.85,4.46){\vector(-1,-4){0.65}} 

\put(4.85,4.32){\vector(-1,-4){0.61}} 

\put(2,1.3){\vector(0,-1){1.1}} 

\put(0.55,2.96){\vector(1,0){1.25}}   

\put(0.55,3.05){\vector(1,1){1.25}}   

\put(0.52,4.46){\vector(1,0){1.28}}   

\thinlines 

\put(1.96,-0.07){\Large I}

\put(4.16,-0.07){\Large V}

\put(4.13,1.46){\Large IV}

\put(3.9,2.89){\Large VI$_\infty$}

\put(1.93,2.89){\Large VI$_0$}

\put(0.1,2.89){\Large VIII}

\put(4.4,4.39){\Large VII$_\infty$}

\put(1.91,4.39){\Large VII$_0$}

\put(0.2,4.39){\Large IX}

\put(1.96,1.46){\Large II}

\linethickness{0.7mm} 

\put(4.3,4.46){\line(-1,0){1.83}}      

\put(3.8,2.96){\line(-1,0){1.45}}      

\end{picture}

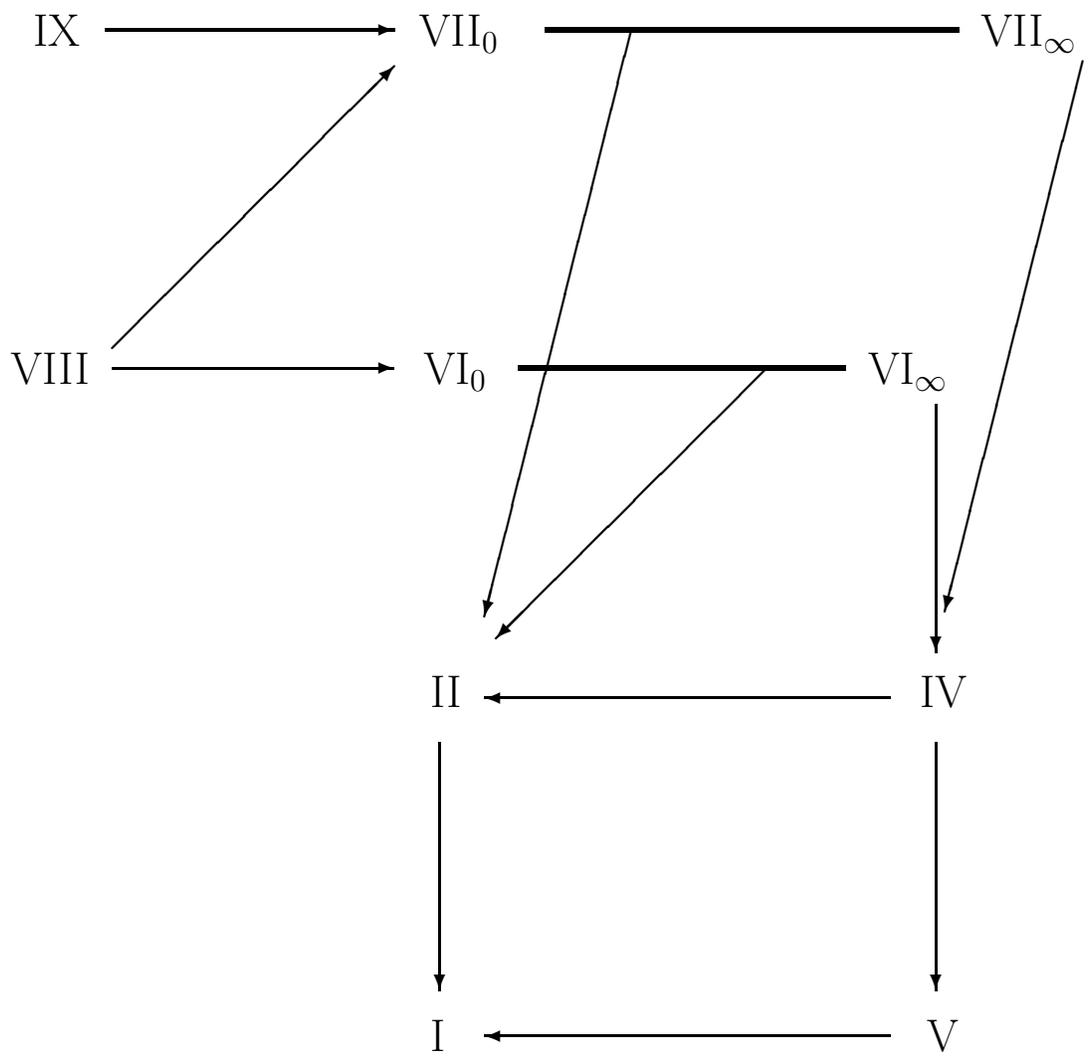
\end{document}